\documentclass[traditabstract]{aa}  
\usepackage{multirow}% http://ctan.org/pkg/multirow
\usepackage{mathrsfs}
\usepackage{longtable}
\usepackage{graphicx} 
\usepackage{txfonts}
\def\kms{\hbox{km\,s$^{-1}$}}
\def\simgr{\,\hbox{\hbox{$ > $}\kern -0.8em \lower 1.0ex\hbox{$\sim$}}\,}
\def\simle{\,\hbox{\hbox{$ < $}\kern -0.8em \lower 1.0ex\hbox{$\sim$}}\,}

\def\o{$\phantom{0}$}
\def\p{$\phantom{-}$}
\def\od{$\phantom{.}$}
\usepackage{color}

\begin{document}

   \title{Mapping the core of the Tarantula Nebula with VLT-MUSE\thanks{Based on observations made with ESO telescopes 
at the Paranal observatory under programme ID 60.A-9351(A).} }
   \subtitle{I. Spectral and nebular content around R136}
   \author{N.~Castro\inst{1}, P.~A.~Crowther\inst{2}, C.~J.~Evans\inst{3}, J.~Mackey\inst{4}, 
N.~Castro-Rodriguez\inst{5,6,7}, J.~S.~Vink\inst{8}, J.~Melnick\inst{9} and F. Selman\inst{9} }
     \institute{Department of Astronomy, University of Michigan, 1085 S. University Avenue, Ann Arbor, MI 48109-1107, 
USA\\
              \email{ncastror@umich.edu}
          \and
            Department of Physics \& Astronomy, University of Sheffield, Hounsfield Road, Sheffield, S3 7RH, UK
         \and
                  UK Astronomy Technology Centre, Royal Observatory, Blackford Hill, Edinburgh, EH9 3HJ, UK
                  \and
                Dublin Institute for Advanced Studies, 31 Fitzwilliam Place, Dublin, Ireland
                \and 
                   GRANTECAN S. A., E-38712, Bre\~na Baja, La Palma, Spain
                   \and
                   Instituto de Astrof\'isica de Canarias, E-38205 La Laguna, Spain 
                   \and
                   Departamento de Astrof\'isica, Universidad de La Laguna, E-38205 La Laguna, Spain
                   \and
               Armagh Observatory and Planetarium, College Hill, Armagh BT61 9DG, Northern Ireland, UK
                  \and
           European Southern Observatory, Alonso de Cordova 3107, Santiago, Chile 
} % \date{Received September 15, 1996; accepted March 16, 1997} \date{Received --; accepted --} 
\titlerunning{Spectral and nebular content around R136 with MUSE
} \authorrunning{N. Castro et al.} % \abstract{1 } % {2} % {3} % {4} % {5} % {6} token are mandatory
 
\abstract{We introduce VLT-MUSE observations of the central
  2$'\times2'$ (30\,$\times$\,30\,pc) of the Tarantula Nebula in the
  Large Magellanic Cloud. The observations provide an unprecedented
  spectroscopic census of the massive stars and ionised gas in the
  vicinity of R136, the young, dense star cluster located in
  NGC\,2070, at the heart of the richest star-forming region in the
  Local Group.  Spectrophotometry and radial-velocity estimates of the
  nebular gas (superimposed on the stellar spectra) are provided for
  2255 point sources extracted from the MUSE datacubes, and we present
  estimates of stellar radial velocities for 270 early-type stars
  (finding an average systemic velocity of 271\,$\pm$\,41\,\kms). We
  present an extinction map constructed from the nebular Balmer lines,
  with electron densities and temperatures estimated from intensity
  ratios of the [\ion{S}{ii}], [\ion{N}{ii}], and [\ion{S}{iii}] lines. The
  interstellar medium, as traced by H$\alpha$ and
  [\ion{N}{ii}]\,$\lambda$6583, provides new insights in regions where
  stars are probably forming. The gas kinematics are complex, but with
  a clear bi-modal, blue- and red-shifted distribution compared to the
  systemic velocity of the gas centred on R136.  Interesting
  point-like sources are also seen in the eastern cavity, western
  shell, and around R136; these might be related to phenomena such
  as runaway stars, jets, formation of new stars, or the interaction
  of the gas with the population of Wolf--Rayet stars.  Closer
  inspection of the core reveals red-shifted material surrounding the
  strongest X-ray sources,
   although we are unable to investigate the kinematics in detail as 
  the stars are spatially unresolved in the MUSE data.  Further papers in
  this series will discuss the detailed stellar content of NGC\,2070
  and its integrated stellar and nebular properties.}

\keywords{Stars: early-types -- Stars: massive --
  open clusters and associations: individual: R136 -- ISM: kinematics
  and dynamics -- ISM: structure -- Galaxies: individual: Large
  Magellanic Cloud}
   \maketitle % %________________________________________________________________

\section{Introduction}

The Tarantula Nebula (30~Doradus) is the most luminous star-forming
complex in the Local Group \citep{1984ApJ...287..116K} and serves as
the closest analogue to the intense star-forming clumps seen in
high-redshift galaxies \citep[e.g.][]{2010MNRAS.404.1247J}. Its
location in the Large Magellanic Cloud (LMC), at a distance of
49.9\,kpc \citep{2013Natur.495...76P}, combined with low foreground
extinction, enables the study of star formation across the full range
of stellar masses, while also revealing the interplay between massive
stars and the interstellar medium (ISM) at sub-parsec scales. Indeed,
30~Dor has been the subject of many studies across the electromagnetic
spectrum, ranging from  X-ray \citep{2006AJ....131.2140T} and $\gamma$-ray  
\citep{2015Sci...347..406H} to optical
\citep[e.g.][]{2011A&A...530A.108E, 2013AJ....146...53S}, infrared
\citep{2015ApJ...807..117Y}, millimetre \citep{2013ApJ...774...73I},
and radio \citep{1978MNRAS.185..263M}.

The Tarantula Nebula extends across several hundred parsec, with star
formation proceeding over the past 15--30\,Myr
\citep{2015A&A...574A..13E, 2016ApJ...833..154C}, as witnessed by the
relatively mature cluster Hodge~301. The star-formation rate has increased more
recently, peaking 1--3\,Myr ago in NGC\,2070
\citep{2015ApJ...811...76C}, the central ionised region that spans
40\,pc and hosts the massive, dense star cluster R136 at its core
\citep[see Table 1 of][]{1991IAUS..148..145W}. Star formation
is indeed still ongoing in NGC\,2070, as evidenced by massive young stellar
objects seen at near-IR wavelengths \citep{1999AJ....117..225W,
  2013AJ....145...98W}.

NGC\,2070 hosts a rich population of well-studied OB-type and Wolf--Rayet
(W--R) stars \citep[e.g.][]{1985A&A...153..235M, 1999A&A...341...98S,
  2011A&A...530A.108E}. Because of the severe crowding, high spatial
resolution observations have been required to investigate the content of
R136, revealing massive O and hydrogen-rich WN stars
\citep{1998ApJ...493..180M} and a young age of 1--2 Myr
\citep{1998ApJ...509..879D,2016MNRAS.458..624C}. In addition, the variety of structures
and cavities in the ISM %carved and blown by stellar winds/supernovae
\citep{1994ApJ...425..720C,2002ApJ...566..302M, 2006AJ....131.2140T}
highlight the signficant feedback from stellar activity in the region
\citep{2002AJ....124.1601W, 2010ApJS..191..160P}.

Given its significance in the context of the formation and evolution of
massive stars, and their interactions with the ISM, NGC\,2070 was an
ideal target for Science Verification (SV) observations with the (then new)
Multi Unit Spectroscopic Explorer \citep[MUSE;][]{2014Msngr.157...13B}
on the Very Large Telescope (VLT). The good spatial sampling (0.2$''$)
and relatively large field (1$'$$\times$$1'$) of MUSE provided a unique
opportunity to characterise the stellar content of NGC\,2070, combined
with estimates of stellar and nebular kinematics. 

\begin{figure*}[] 
\resizebox{\hsize}{!}{\includegraphics[angle=0,width=\textwidth]{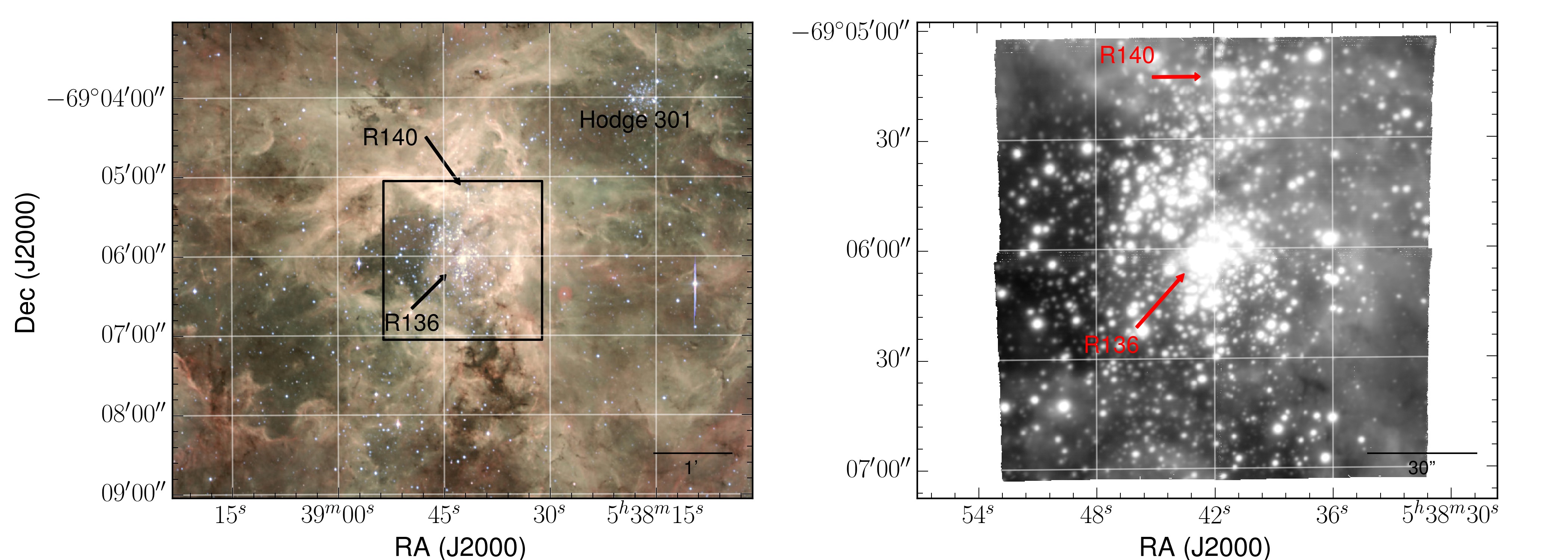}}
  \caption{{\it Left:} Total 2$'$\,$\times$2\,$'$ field observed with
    MUSE (black square) overlaid on a colour-composite image of the
    central 7.3$'$\,$\times$\,6$'$ ($\approx$110\,$\times$\,90\,pc) of
    the Tarantula Nebula obtained with the Wide Field Imager (WFI).
    The young star cluster R136 is indicated at the centre, with R140,
    an aggregate of W--R stars, to the north and 
    Hodge\,301, the older cluster, to the north-west. {\it Right:}
    Continuum-integrated (4600-9300\,\AA) MUSE mosaic.}
  \label{Fig:IMAG} \end{figure*}

This article presents the MUSE SV observations of NGC\,2070 and is
structured as follows. Sect.~\ref{OBS} gives an overview of the data
and Sect.~\ref{PHO} briefly summarises the stellar content.  The
characteristics of the ionised gas from the MUSE datacubes are
described in Sect.~\ref{NEB}, including maps of extinction and
electron density/temperature from the ratios of nebular emission
lines.  Sect.~\ref{VELO} explores the stellar and gas kinematics of
the region, extending previous ISM studies
\citep[e.g.][]{1994ApJ...425..720C,2002AJ....124.1601W,2010ApJS..191..160P,2013A&A...550A.108V,2017MNRAS.469.3424M},
with brief conclusions in Sect.~\ref{CONC}. Future papers will focus on
the stellar content of NGC\,2070 and its integrated stellar and
nebular properties.

\section{Observations} \label{OBS}

NGC\,2070 was observed as part of the MUSE SV programme at the VLT in
August 2014. Four overlapping fields, each with an individual field of
1$'$$\times$1$'$ and a pixel scale of 0.2$''$/pixel, were observed with
4$\times$10\,s, 4$\times$60\,s, and 4$\times$600\,s exposures.
Table~\ref{tab:obs} summarises the observations, central pointings,
and average point-spread functions (PSFs) for the four fields.

The data were reduced using the MUSE pipeline based on {\sc esorex}
recipes \citep{2012SPIE.8451E..0BW}, with final astrometric
calibration undertaken with the catalogue of
\citet{1999A&A...341...98S}. The exposures were combined to increase
the signal-to-noise ratio (S/N) and to mitigate the effects of cosmic
rays and instrumental features (i.e. the patterns of the MUSE
image-slicers and spectrographs). The final spectra span
4595-9366\,\AA, with a resolving power of $R$\,$\approx$3000
around H$\alpha$. Relative flux calibration is achieved using
observations of a spectrophotometric standard each night, but
absolute fluxes require additional calibration (see Sect.~\ref{PHO}).

The 2$'$$\times$2$'$ mosaic observed with MUSE is shown in
Fig.~\ref{Fig:IMAG} on a colour-composite of the region from the Wide Field Imager
(WFI) on the 2.2m MPG/ESO telescope (La Silla)\footnote{Observations 
obtained under programme 076.C-0888, processed and released by the ESO VOS/ADP group.}.
The MUSE field takes in much of NGC\,2070, centred on R136 and including
R140 to the north (but excluding Hodge\,301, the older cluster to the north-west).

\begin{table}[h]
        \caption[]{Observing log and average point-spread function (PSF).}
        \small
        \label{tab:obs}

                        \begin{tabular}{lccccc}
                                \hline
                                \hline
                                Field & Date & $\alpha$ [${\rm h,m,s}$] & $\delta$ [${^\circ,',''}]$ & PSF [$''$] \\
                                \hline
                                Field A & 2014 Aug 18 & 5 38 36.7  & $-$69 06 33.98 & 0.7 \\
                                Field B & 2014 Aug 21 & 5 38 47.4  & $-$69 06 33.98 & 0.9 \\
                                Field C & 2014 Aug 23 & 5 38 36.7  & $-$69 05 35.99 & 1.1 \\
                                Field D & 2014 Aug 20 & 5 38 47.4  & $-$69 05 35.99 & 1.0 \\
                                \hline
                        \end{tabular} 
\end{table}

\section{Stellar content of NGC\,2070} \label{PHO}

We extracted spectra from the reduced MUSE cubes using detections of
2255 sources from SExtractor \citep{1996A&AS..117..393B} on the
continuum-integrated mosaic shown in the right-hand panel of
Fig.~\ref{Fig:IMAG} (constructed using the 600\,s frames).  The
extracted MUSE sources are listed (in ascending right ascension) in
Table~\ref{TAB:cat}, where Cols. 2 and 3 give previous identifications
from \citet{1999A&A...341...98S} and \citet{2011A&A...530A.108E},
respectively. Given the resolved ISM structures and stellar crowding
in the region, we estimate the sky contribution locally to each
source, adopting annuli with inner radii of 7 pixels and a radial
width of 1 pixel. Even with this approach, the complex nebulosity and
nearby stars still thwart ideal sky subtraction in many cases.

We were relatively conservative in the extraction of sources, in the
sense that there were several hundred additional sources detected by
SExtractor  where the MUSE magnitudes and spectra were
probably contaminated by nearby stars. We
note that R136 was not explicitly excluded from the SExtractor
analysis, but that very few sources were detected because of the
significant crowding.

\begin{figure*}[]
\resizebox{\hsize}{!}{\includegraphics[angle=0,width=\textwidth]{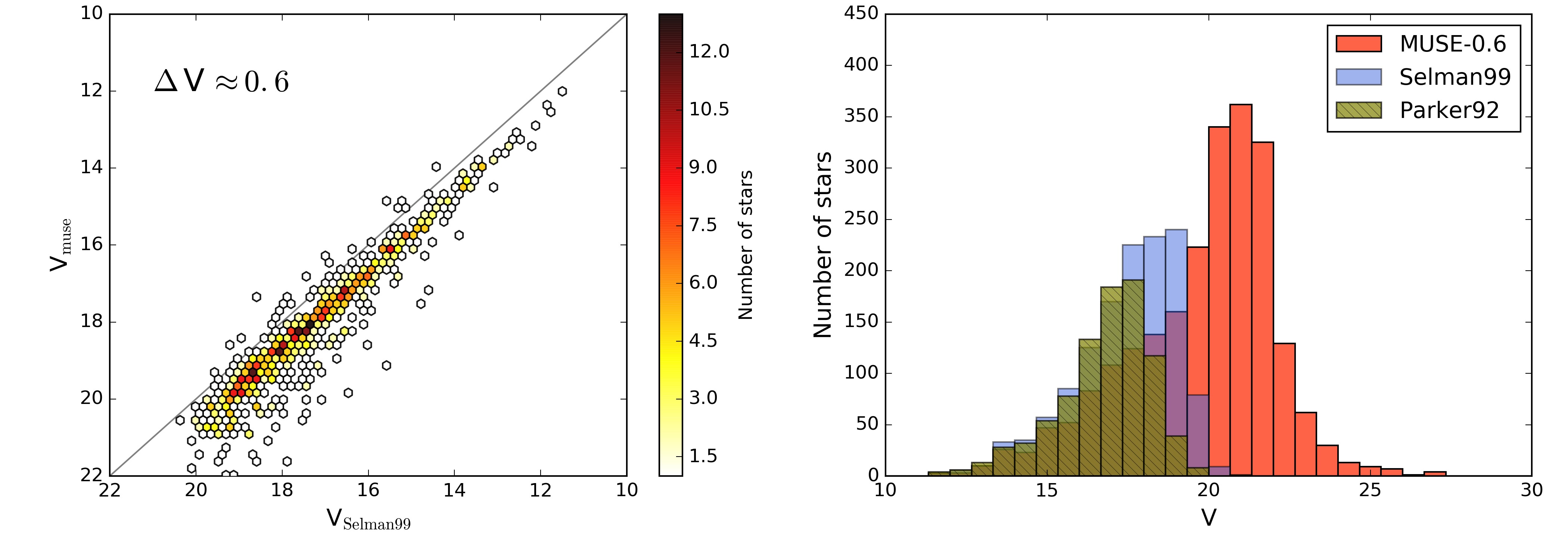}}
\caption{{\it Left:} Calibration of MUSE $V$-band magnitudes compared with photometry 
	from \cite{1999A&A...341...98S}; the solid black line
  indicates the 1:1 ratio. {\it Right:} Magnitude distribution of
  stars extracted from the MUSE data (and shifted by $-$0.6\,mag to
  match \citeauthor{1999A&A...341...98S}); the corresponding distributions
  for the MUSE area from \cite{1992PASP..104.1107P} and
  \cite{1999A&A...341...98S} are shown for comparison.}
        \label{Fig:PHOTO_SEL} 
\end{figure*}

\subsection{Photometry}\label{phot}

Johnson $V$- and $I$-band photometry is obtained for each MUSE source
using the relative fluxes from the pipeline.  To estimate the $V$-band
zero-point, we compared the instrumental MUSE magnitudes with published
photometry for matched sources from \citet{1999A&A...341...98S}, as
shown in the left-hand panel of Fig.~\ref{Fig:PHOTO_SEL}. A zero-point
correction of $-$0.6\,mag is necessary to shift the MUSE photometry to
the \citet{1999A&A...341...98S} data \citep{2015A&A...582A.114W}. The scatter at faint magnitudes
probably arises from issues with background subtraction, that
is, contamination by nearby stars (particularly around R136).
\citet{1999A&A...341...98S} observed NGC\,2070 with $UBV$ filters, and in
absence of comparable $I$-band data, we therefore estimated ($V-I$)$_{\rm MUSE}$
on the basis of the relative flux calibration from the
spectrophotometric standards observed with MUSE.

The estimated MUSE magnitudes and colours are included in
Table~\ref{TAB:cat}; 1147 sources have $V_{\rm MUSE}$\,$<$\,21.5. The
right-hand panel of Fig.~\ref{Fig:PHOTO_SEL} compares the distribution
of $V$-band magnitudes of the MUSE sources to ground-based photometry
from \citet{1992PASP..104.1107P} and \citet{1999A&A...341...98S}. The
MUSE observations go deeper, but are more limited by crowding at the
brighter end. This is not surprising given  the typical PSF of the
MUSE data; see the $\approx$0.7$''$ obtained by
\citet{1999A&A...341...98S} in the $V$ band\footnote{We acknowledge
  that higher-resolution imaging of the region is available from the Hubble
  Tarantula Treasury Project
  \citep[HTTP;][]{2013AJ....146...53S,2016ApJS..222...11S}. Given the
  signficant difference in image quality, cross-matching between these
  data and the MUSE catalogue is non-trivial, often with multiple (and
  sometimes spurious) matches. Our primary interest here are the
  extracted MUSE spectra, therefore we did not employ the HTTP data further,
  but we caution that some of our sources will be
  multiples or composites if observed at finer angular resolution.}.

\begin{figure}[]

\resizebox{\hsize}{!}{\includegraphics[angle=0,width=\textwidth]{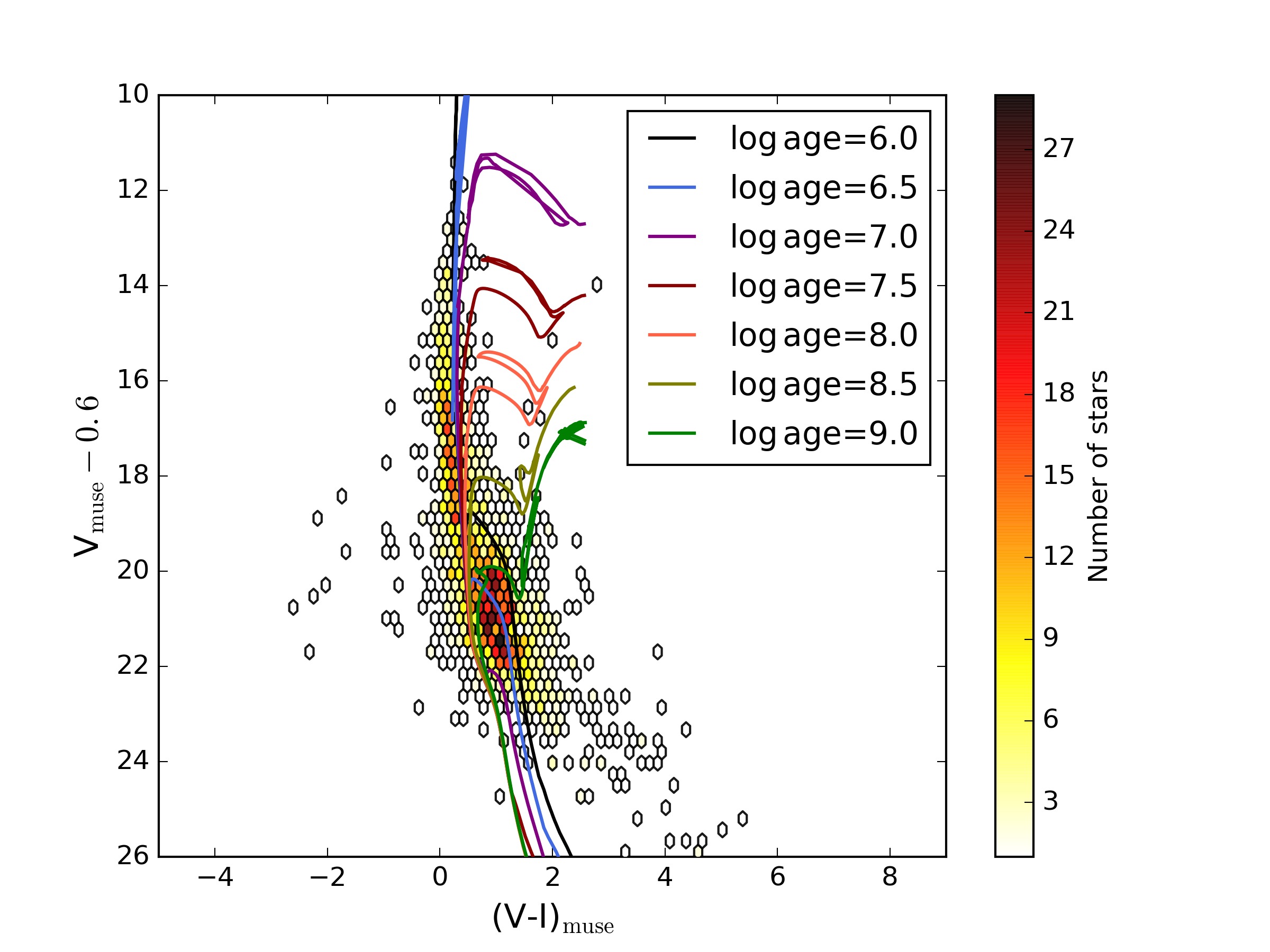}}
        
  \caption{Colour-magnitude diagram for the extracted MUSE sources
    compared with isochrones from {\sc parsec} evolutionary models
    \citep{2012MNRAS.427..127B}; the ages quoted are in years.}
        \label{Fig:VI} \end{figure}

The locations of our extracted MUSE sources in a colour-magnitude
diagram are shown in Fig.~\ref{Fig:VI} compared to isochrones spanning
1\,Myr to 1\,Gyr from {\sc parsec} stellar evolution models for the
metallicity of the LMC, which include both pre-main-sequence and
main-sequence phases
\citep{2012MNRAS.427..127B}\footnote{http://sted.oapd.inaf.it/cgi-bin/cmd}.
As expected, Fig.~\ref{Fig:VI} illustrates the significant population
of luminous blue stars that is compatible with ages of a few Myr.
Figure~\ref{Fig:VI} is also consistent with a significant
pre-main-sequence population (with 20\,$\leq$\,V\,$\leq$\,22 and
($V-I$)\,$\approx$1) with ages of a few Myr \citep[see
e.g.][]{2015ApJ...811...76C,2016ApJS..222...11S,2017arXiv170302876K}.
Figure~\ref{Fig:VI} does not allow distinguishing multiple young
bursts \citep{1997ApJS..112..457W,1998ApJ...493..180M}, although older
populations are also present in NGC\,2070 (e.g. a few cool supergiants
with ages of $\approx$30 Myr). We note that \cite{2016ApJS..222...11S}
reported evidence for an older underlying stellar population
\citep[see also][]{2009AJ....138.1243H}. This is part of the local
field population of the LMC and could influence the analysis of the
pre-main-sequence stars.

\begin{figure*}[]
\resizebox{\hsize}{!}{\includegraphics[angle=0,width=\textwidth]{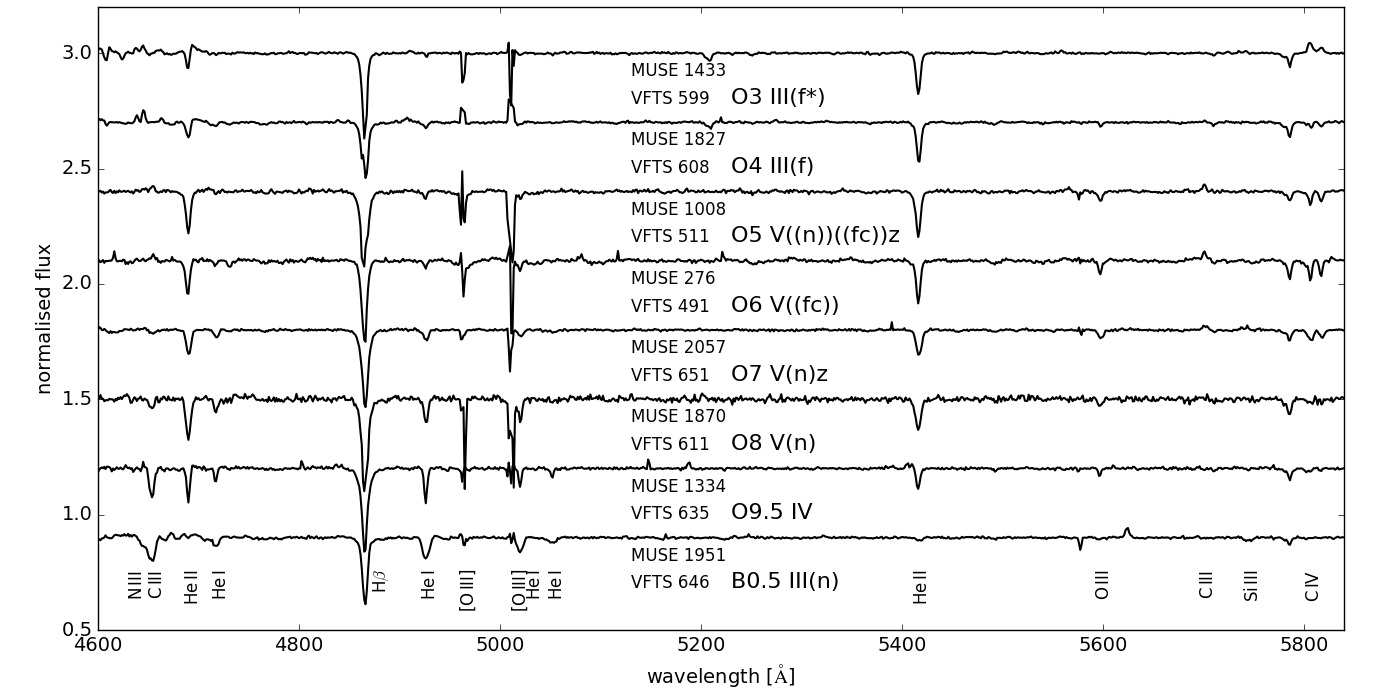}}
  \caption{Illustrative O- and early B-type spectra from the MUSE
    observations, with classifications from
    \citet{2014A&A...564A..40W} and \citet{2015A&A...574A..13E}. The relevant lines for
spectral classification are labelled, as is the location of the nebular [\ion{O}{iii}] lines (which show
residuals from over- and under-subtraction of the local nebulosity).}
        \label{Fig:spec} 
\end{figure*}

\begin{figure}[]
\resizebox{\hsize}{!}{\includegraphics[angle=0,width=\textwidth]{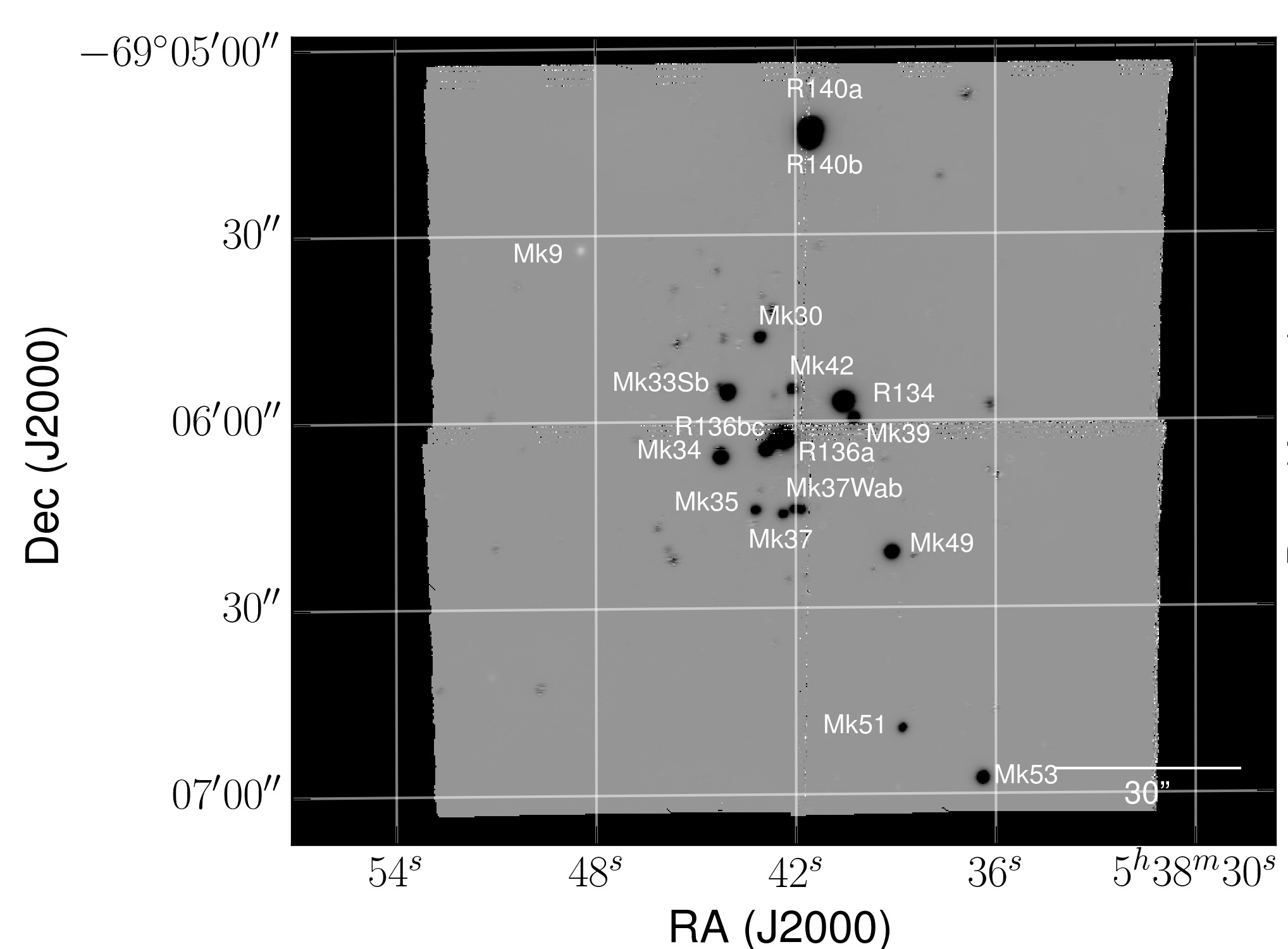}}
  \caption{Net \ion{He}{ii}\,$\lambda$4686 emission in the MUSE mosaic
of NGC\,2070, highlighting the Wolf--Rayet content of R136 and R140,
    and other emission-line sources. The source north-east of R136
,    which is brighter in the continuum (cf. $\lambda$4686), is the M-type
supergiant Mk\,9.}\label{Fig:4686} 
\end{figure}

\subsection{Spectroscopy}

The MUSE observations provide intermediate-resolution spectroscopy at
S/N$\geq$50 for 588 stars (from the 4\,$\times$\,600\,s exposures).
The observed wavelength range includes the primary classification
diagnostics of Of/WN and W--R stars (except for \ion{N}{iv} 4058), but not
the features used for the majority of early-type stars
\citep[e.g.][]{1990PASP..102..379W,2008A&A...485...41C,2012A&A...542A..79C}.
Nonetheless, the MUSE data are expected to permit robust classifications through
alternative diagnostics. For O-type stars the spectra
include several \ion{He}{i} ($\lambda\lambda$4713, 4921, 5876, 6678)
and \ion{He}{ii} ($\lambda\lambda$5412, 6683) transitions, as
well as \ion{N}{iii} $\lambda\lambda$4634-41, \ion{N}{iv}
$\lambda\lambda$7103-29, \ion{N}{v} $\lambda\lambda$4603-20,
\ion{C}{iii}\,$\lambda\lambda$4647-51, $\lambda$5696, \ion{C}{iv}
$\lambda\lambda$5801-12, and \ion{O}{iii}\,$\lambda$5592. The H$\beta$
and H$\alpha$ lines are also included, but are often severely affected
by nebulosity and issues related to sky substraction. For B-type
stars there are additional useful \ion{He}{i} lines, as well
as \ion{N}{ii}
$\lambda\lambda$4601-43, $\lambda\lambda$5667-5701, \ion{O}{ii}
$\lambda\lambda$4639-76, and \ion{Si}{iii}\,$\lambda$5740.

\begin{table*}
  \begin{center}
    {\small
      \caption[]{Dominant stellar emission-line sources within the
        MUSE observations of NGC\,2070, with identifications from
        \citet[][BAT99]{1999A&AS..137..117B}, \citet[][R]{1960MNRAS.121..337F}, 
        \citet[][Mk]{1985A&A...153..235M},
        \citet[][VFTS,]{2011A&A...530A.108E} and the MUSE data. }\label{TAB:4686}
\begin{tabular}{rrrrrrcp{2.1cm}p{1.7cm}p{1.7cm}p{1.7cm}}
\hline\hline
BAT99 & R & Mk & VFTS & MUSE & Spectral type & Ref. & $F_{\ion{N}{iii} \lambda 4640+\ion{C}{iii} \lambda 4650}$ &
$F_{\ion{He}{ii} \lambda 4686}$ & $F_{\ion{C}{iv} \lambda 5808}$ & $F_{H\alpha}$ \\
      &       &      &         &         &         &     &  [$10^{-13}$] & [$10^{-13}$] & [$10^{-13}$] & [$10^{-13}$] \\
\hline
96      & -- & 53     & 427    &  389 & WN8(h)       & a & \o\o2.0\o\,$\pm$\,0.1    & \o4.9\,$\pm$\,0.1 & \o\o0.17\,$\pm$\,0.06 & \o3.0\,$\pm$\,0.1 \\
97      & -- & 51     & 457    &  603 & O3.5\,If/WN7 & b & \o\o0.43\,$\pm$\,0.03    & \o1.3\,$\pm$\,0.1 & \o\o0.07\,$\pm$\,0.01 & \o1.7\,$\pm$\,0.1\\
98      & -- & 49     & -- & 1261 & WN6(h)       & a & \o\o2.2\o\,$\pm$\,0.1    & 11.4\,$\pm$\,0.1  & \o\o0.55\,$\pm$\,0.10 & \o8.6\,$\pm$\,0.1\\
99      & -- & 39     & 482    & 2003 & O2.5\,If/WN6 & b & \o\o0.80\,$\pm$\,0.09    & \o2.1\,$\pm$\,0.1 & \o\o0.21\,$\pm$\,0.02 & \o2.3\,$\pm$\,0.1\\
100     & 134    & -- & 1001   & 1978 & WN6h         & a & \o\o6.3\o\,$\pm$\,0.3    & 21.6\,$\pm$\,0.2  & \o\o0.83\,$\pm$\,0.08 & \o9.3\,$\pm$\,0.1\\
101-102 & 140a   & -- & 507    & 3191 & WC5+WN6+O    & a & \o\o115\od\,$\pm$\,2      & 19.1\,$\pm$\,0.6  & \o\o110\od\,$\pm$\,1   & 10.2\,$\pm$\,0.2 \\
103     & 140b   & -- & 509    & 3174   & WN6        & a & \o\o3.4\o\,$\pm$\,0.2    & 20.9\,$\pm$\,0.2  & \o\o-- & \o6.9\,$\pm$\,0.3\\
104     & -- & 37Wb   & -- & 1374   & O2\,If/WN5 & b & \o\o0.30\,$\pm$\,0.06    & \o1.6\,$\pm$\,0.1   & \o\o0.05\,$\pm$\,0.02 & \o1.1\,$\pm$\,0.1\\
--  & -- & 37Wa   & -- & 1349   & O4\,If$^+$ & d & \o\o0.51\,$\pm$\,0.05    & \o1.2\,$\pm$\,0.1   & \o\o--            & \o1.2\,$\pm$\,0.1\\
105     & -- & 42     & -- & 2102   & O2\,If     & c & \o\o0.54\,$\pm$\,0.09    & \o1.9\,$\pm$\,0.1   & \o\o0.09\,$\pm$\,0.02 & \o1.1\,$\pm$\,0.1\\
106-110 & 136a   & -- & -- & -- & WN5h+O     & a & \o\o--    & 30.0\,$\pm$\,0.1   & \o\o-- & \p--\\ 
--  & -- & 37     & -- & 1442   & O4\,If$^+$ & d & \o\o0.70\,$\pm$\,0.08    & \o1.4\,$\pm$\,0.1   & \o\o0.21\,$\pm$\,0.01 & \o2.0\,$\pm$\,0.1\\
111     & 136b   & -- & -- & 1669   & O4\,If/WN8 & e & \o\o0.57\,$\pm$\,0.10    & \o4.0\,$\pm$\,0.1   & \o\o0.11\,$\pm$\,0.03 & \o3.1\,$\pm$\,0.1\\
112     & 136c   & -- & -- & 1737   & WN5h       & a & \o\o0.87\,$\pm$\,0.30    & \o6.6\,$\pm$\,0.1   & \o\o0.28\,$\pm$\,0.04 & \o3.8\,$\pm$\,0.1\\
113     & -- & 30     & 542    & 2999   & O2\,If/WN5 & b & \o\o0.39\,$\pm$\,0.08    & \o2.6\,$\pm$\,0.1   & \o\o0.13\,$\pm$\,0.01 & \o1.8\,$\pm$\,0.1\\
114     & -- & 35     & 545    & 1474   & O2\,If/WN5 & b & \o\o--               &  \o1.6\,$\pm$\,0.1   & \o\o0.17\,$\pm$\,0.01 & \o1.4\,$\pm$\,0.1\\
115     & -- & 33Sb   & -- & 2177   & WC5        & a & \o\o13.6\,$\pm$\,0.5     & \o1.9:              & \o\o8.4\o\,$\pm$\,0.1 & \o--          \\
116     & -- & 34     & -- & 1766   & WN5h       & a & \o\o0.51\,$\pm$\,0.08    & 10.3\,$\pm$\,0.1 & \o\o0.45\,$\pm$\,0.05 & \o7.4\,$\pm$\,0.1\\
\hline
\end{tabular} 
\tablefoot{Observed fluxes are given in cgs units (erg\,s$^{-1}$\,cm$^{-2}$). Refs: (a) \citet{1999A&AS..137..117B}; (b) \citet{2011MNRAS.416.1311C}; (c) \citet{2002AJ....123.2754W};
(d) \citet{1998ApJ...493..180M}; (e) \citet{2016MNRAS.458..624C}.
}
}
\end{center}
\end{table*}

Illustrative MUSE spectra over the range 4600-6000\,\AA\ are shown in
Fig.~\ref{Fig:spec} for eight O- and early B-type stars previously
classified by \cite{2014A&A...564A..40W} and
\cite{2015A&A...574A..13E}; the relevant diagnostic lines are
indicated. We note that the sky subtraction can give residuals in the
[\ion{O}{iii}]\,$\lambda\lambda$4959, 5007 lines, and can similarly
influence the appearance of H$\beta$. Classification and quantitative
analysis of this large dataset of early-type stars requires
implementation of automated techniques, and results on this aspect of
the MUSE data will be reported in a separate article.

It is well established that NGC\,2070 contains a number of early-type
stars with strong emission lines \citep{1985A&A...153..235M}.
Figure~\ref{Fig:4686} shows a net \ion{He}{ii}\,$\lambda$4686 image from
the MUSE data, obtained by subtracting the local continuum
($\lambda\lambda$4740--60) from the emission between
$\lambda\lambda$4680-4700. Each of R136a (containing several WN and
Of/WN stars), R140a (WN+WC), R140b (WN), and R134 (WN6) are prominent,
together with several WN stars (Mk\,34, R136c, Mk\,49, Mk\,53), Of/WN
stars (Mk\,39, Mk\,51, R136b), Of stars (Mk\,42, Mk\,37), and a WC star
(Mk33Sb). The observed emission-line fluxes of
\ion{He}{ii}\,$\lambda$4686,
\ion{N}{iii}\,$\lambda$4640+\ion{C}{iii}\,$\lambda$4650,
\ion{C}{iv}\,$\lambda$5808, and H$\alpha$ for these sources are
presented in Table~\ref{TAB:4686}, illustrating that several W--R
stars contribute to the cumulative $\lambda$4686 emission, with two WC
stars (R140a1, Mk33Sb) dominating the $\lambda$4640--50 emission.
From across the entire NGC\,2070 region, R140a1 provides the
dominant contribution to both the blue W--R bump and the \ion{C}{iv}
$\lambda\lambda$5801-12 W--R feature.

\begin{figure*}[]
\resizebox{\hsize}{!}{\includegraphics[angle=0,width=\textwidth]{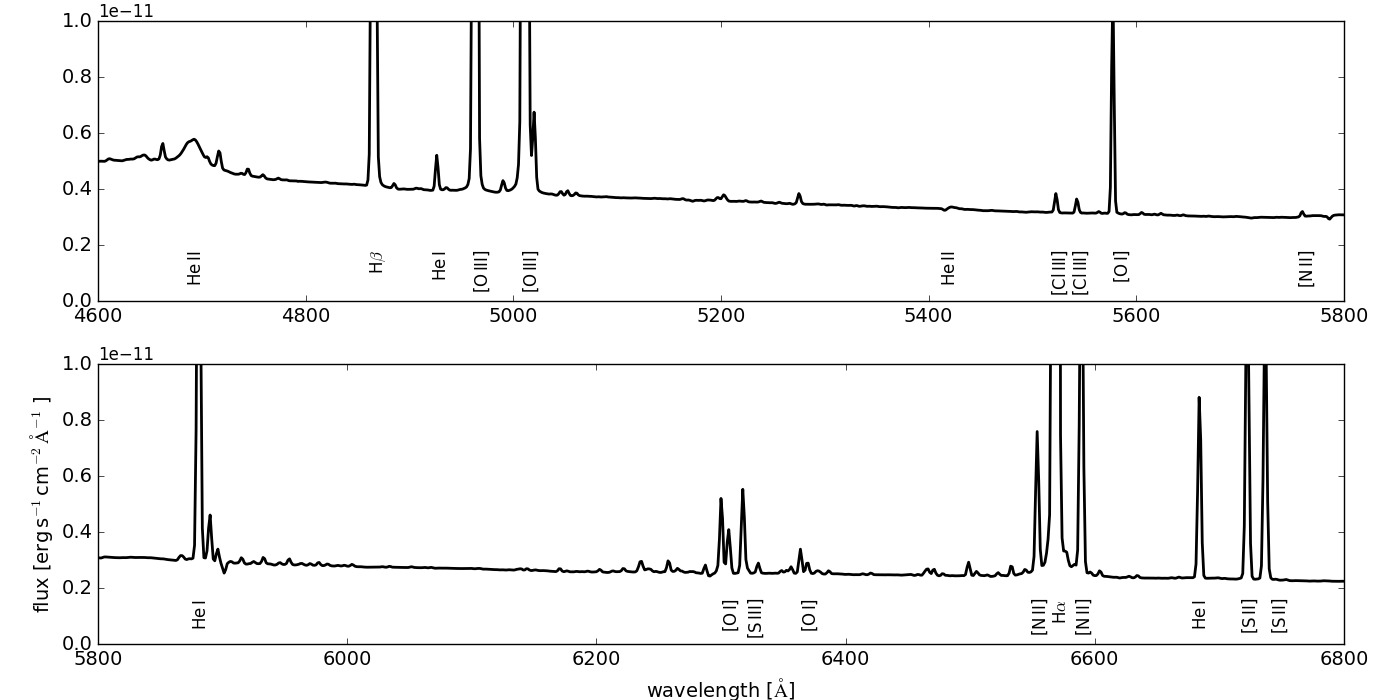}}
        \caption{Integrated spectrum (4600-6800\,\AA) from the MUSE mosaic; prominent emission lines are labelled.}
        \label{Fig:spec_neb} 
\end{figure*}

\section{Ionised gas in NGC\,2070} \label{NEB}

The MUSE data also provide a wealth of spatially resolved information
on the nebular content of NGC\,2070.  The integrated spectrum
from the MUSE cubes is shown in Fig.~\ref{Fig:spec_neb}, displaying
very strong H$\alpha$, H$\beta,$ and
[\ion{O}{iii}]\,$\lambda\lambda$4959, 5007 emission, and with
prominent emission from \ion{He}{i}\,$\lambda$5876, 6678,
[\ion{N}{ii}]\,$\lambda\lambda$6548-83, and
[\ion{S}{ii}]\,$\lambda\lambda$6717, 6731.   Broad emission is 
present around \ion{He}{ii}\,$\lambda$4686 as a result of
the strong contribution from the population of W--R stars in NGC\,2070 (see Fig.~\ref{Fig:4686}).

We now discuss the spatial distribution of the ionised gas in
NGC\,2070, based on the low-ionisation diagnostic lines (H$\alpha$ and
[\ion{N}{ii}]\,$\lambda$ $6583.45$, hereafter [\ion{N}{ii}]); the gas
kinematics are discussed in Sect.~\ref{VELO}. The H$\alpha$
emission was saturated in large areas of the 600\,s exposures, therefore maps
were created using single-component Gaussian fits to the
4\,$\times$\,60\,s exposures. This gave S/N\,$>$\,50/pix. across the
entire MUSE mosaic for both lines. From similar Gaussian fits to the
relevant emission lines we also constructed maps of the interstellar
extinction, electron density, and temperature using standard nebular
techniques.

\subsection{ Intensity of H$\alpha$ and [\ion{N}{ii}] emission}\label{EMI}

The H$\alpha$ and [\ion{N}{ii}] emission in the MUSE mosaic reveal the
ionised gas and structure of the ISM in NGC\,2070, as shown in the
upper panels of Fig.~\ref{Fig:GAUS_INT} \citep[see
also][]{2002AJ....124.1601W}.  These include a $\approx$1'$\times$30$''$
($\simeq$15\,$\times$\,7\,pc) cavity to the east of R136  and a
  large shell to the west with an angular size of 1$'$
  ($\simeq$15\,pc).  In the north-eastern part of the images is the
  so-called \textit{\textup{eastern filament}}, which includes {\it Knot~1}
  \citep[see][]{1997ApJS..112..457W,1999AJ....117..225W}.

  The [\ion{N}{ii}] flux is significantly lower than that from
  H$\alpha$, but the S/N is sufficient for robust detection across the
  entire MUSE mosaic (except for some regions, e.g. around R140, where
  strong H$\alpha$ emission masks the [\ion{N}{ii}]). The
  [\ion{N}{ii}] emission mimics the structures seen in H$\alpha$, but
  appears to be in more discrete clumps. These are generally
  associated with individual sources, such as W-R
    stars within the cluster core
    (Fig.~\ref{Fig:4686}).  The [\ion{N}{ii}] map reveals many individual
  clumps in the various ISM filaments, which probably trace current
  stellar nurseries \citep[as suggested by][]{1997ApJS..112..457W}.  We
  note two isolated regions in the south-eastern part of the
  [\ion{N}{ii}] map with remarkable bullet-shaped morphologies\footnote{$\alpha$\,$\approx$\,05$^{\rm h}$38$^{\rm
                m}$50$^{\rm s}$, $\delta$\,$\approx$\,$-$69$^\circ$06$'$39$''$}, perhaps
  indicative of runaway phenomena. 
\subsection{ FWHM of H$\alpha$ and [\ion{N}{ii}] emission}\label{SIG}

\begin{figure*}[]
\resizebox{\hsize}{!}{\includegraphics[angle=0,width=\textwidth]{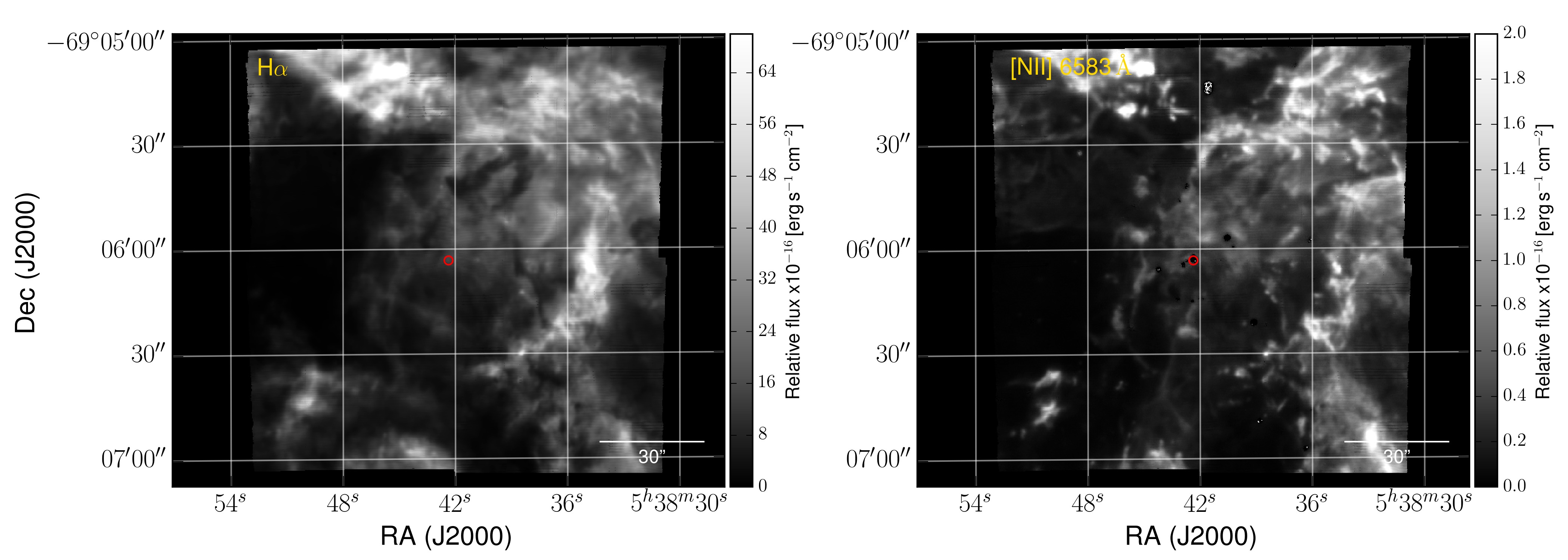}}
\resizebox{\hsize}{!}{\includegraphics[angle=0,width=\textwidth]{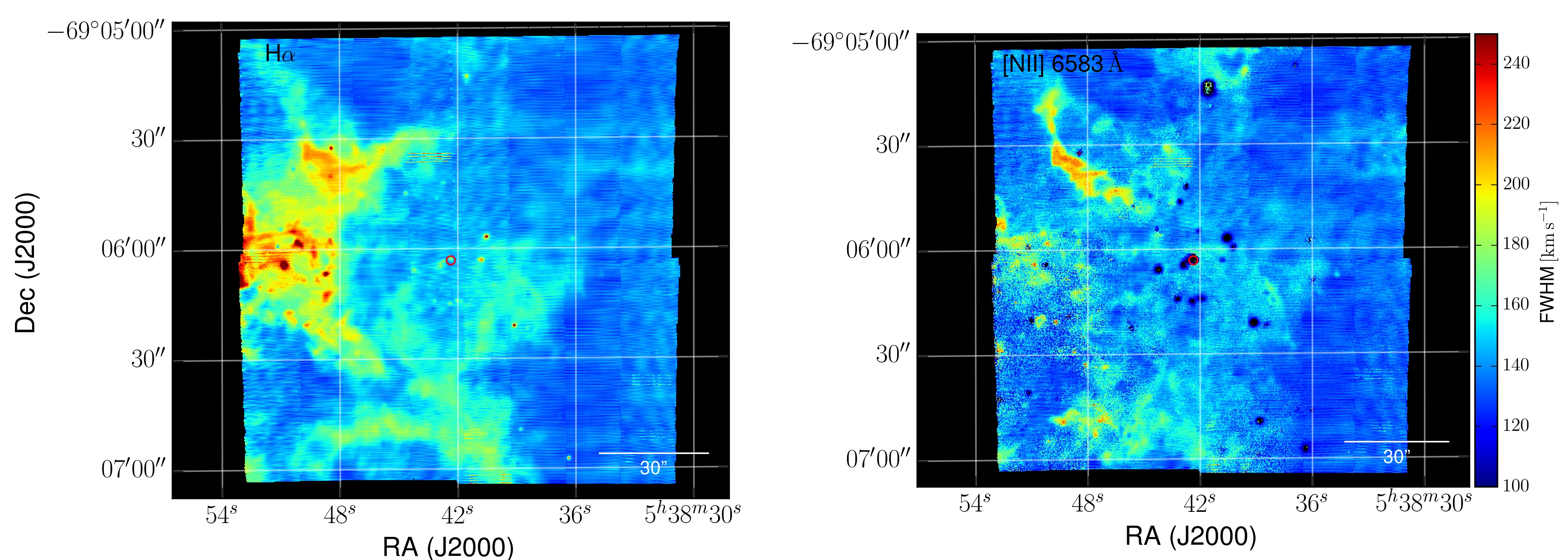}}

\caption{Relative H$\alpha$ (left) and
  [\ion{N}{ii}]\,$\lambda$$6583.45$ (right) fluxes (upper panels) and
  full-width at half-maxima (lower panels) extracted from
  single-component Gaussian fits to both lines in the MUSE datacubes.
  The red circle indicates the core of R136.}\label{Fig:GAUS_INT} \end{figure*}

The lower panels of Fig.~\ref{Fig:GAUS_INT} present the full-width at
half-maxima (FWHM) of single-component Gaussian fits to both the
H$\alpha$ and [\ion{N}{ii}] lines.  The MUSE data appear to reveal a
series of seemingly broad components in the eastern cavity, although
we caution that the velocity resolution of MUSE at H$\alpha$ is 
$\approx$100\,\kms, which means that multiple velocity components may be unresolved.
Past observations of NGC\,2070 at higher spectral resolution
have indeed revealed rich sub-structures
\cite[e.g.][]{1999MNRAS.302..677M,2013A&A...555A..60T}. The
multi-component structures seen previously are outside the MUSE
mosaic, but are close to the eastern edge of our data where the FWHM
shows the highest values. Moreover, the multi-component composition of
the ISM in the cavity was recently reported by
\cite{2017MNRAS.469.3424M} from Fabry-Perot observations.

The large FWHM in Fig.~\ref{Fig:GAUS_INT} also appears to trace the
border between R136 and the cavity, potentially a consequence of the
radiation pressure from massive stars in R136
\citep{2006AJ....131.2140T,2011ApJ...738...34P}. In contrast, the
distribution of FWHM estimates in the west (sampling the large shell)
are more comparable with the velocity resolution of MUSE.

Figure~\ref{Fig:GAUS_INT} indicates several compact regions (in the
eastern cavity and elsewhere) with high FWHM estimates ($\geq$ 250
km\,s$^{-1}$). These are probably associated with stars with dense,
complex circumstellar material, and were also reported by
\citet{2017MNRAS.469.3424M}. For example,
Mk\,9\footnote{$\alpha$\,$=$\,05$^{\rm h}$38$^{\rm m}$48\fs480,
  $\delta$\,$=$\,$-$69$^\circ$05$'$32\farcs58
  \citep{1985A&A...153..235M,2003yCat.2246....0C}.}  (MUSE\,2658), the
red supergiant in the northern part of the cavity, shows a large FWHM
in the H$\alpha$ map, possibly the externally photoionised wind \citep{2015A&A...582A..24M}. A number of the W--R
stars in the mosaic also appear to have high FWHM values, which arise
from the contribution of broad H$\alpha$ emission from their stellar
winds to the nebular component.

For completeness, we note that the Tarantula Nebula has been
  imaged with near-IR narrow-band filters (H$_2$ 2.12\,$\mu$m and
  Br\,$\gamma$\,2.17\,$\mu$m) with the NEWFIRM camera on the 4m
  Blanco telescope on Cerro Tololo \citep{2015ApJ...807..117Y}.
  Inspection of these images (kindly provided by the authors)
  revealed morphological features or trends similar to those discussed
  above.

\subsection{Extinction map} \label{SECT:EXT}

\begin{figure*}[]
\resizebox{\hsize}{!}{\includegraphics[angle=0,width=2\textwidth]{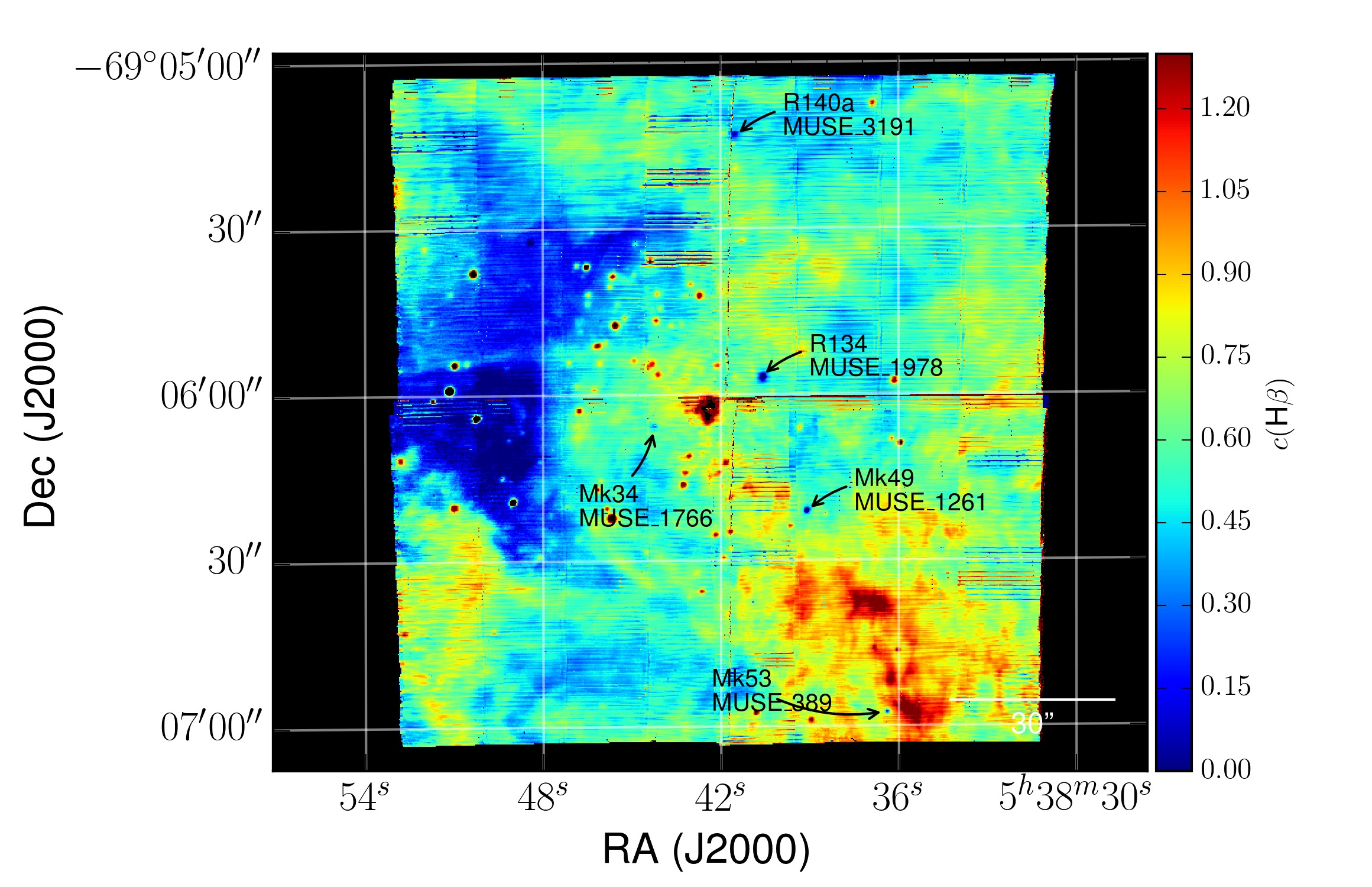}}
\caption{Reddening coefficient $c(H\beta)$ across NGC\,2070. Five W--R
  stars with low $c(H\beta$) values are labelled (see
  Sect.~\ref{SECT:EXT} and Table~\ref{TAB:4686}).}\label{Fig:EXT} 
\end{figure*}

The strong nebular emission in NGC\,2070 permits estimates of
interstellar extinction from the H$\alpha$/H$\beta$ ratio, assuming
Case~B recombination theory for $n_{e}$\,=\,100 cm$^{-3}$ and 
$T_{e}$\,=\,10000\,K, that is,  an intrinsic ratio of
I(H$\alpha$)/I(H$\beta$)\,$=$\,2.86 \citep{1987MNRAS.224..801H}. We
adopted a standard extinction law \citep{1989ApJ...345..245C} with
$R_V$\,$=$\,3.1 in our calculations to ensure a direct comparison with
\cite{2010ApJS..191..160P}.  Fig.~\ref{Fig:EXT} shows $c(H\beta$)
across the MUSE mosaic, with a broad range of extinctions
(0.15\,$\leq$\,c(H$\beta$)\,$\leq$\,1.2), and where foreground
extinction due to the Milky Way is $c$(H$\beta$)\,$\approx$\,0.1. The
average $c$(H$\beta$) of 0.55\,mag equates to E($B-V$)\,$\approx$\,0.38
($\simeq$\,0.7\,$c$(H$\beta$)), in excellent agreement with
\citet{2010ApJS..191..160P}.

The extinction map in Fig.~\ref{Fig:EXT} shows a complex distribution,
but resembling the large ISM clouds highlighted by the H$\alpha$
intensity map (Fig.~\ref{Fig:GAUS_INT}), for example,  lower extinction in the
eastern cavity.  Several regions of high extinction are seen to the
south-west, but with no obvious counterparts in the H$\alpha$ map.  The
low extinction toward several of the W--R stars in the region
(relative to their local environment, see sources labelled in
Fig.~\ref{Fig:EXT}) appears remarkable. This suggests that their stellar
winds and/or ionising radiation could have influenced their immediate
ISM. While the extinction determinations near these W--R stars may be
influenced by extended stellar emission at H$\alpha$ and H$\beta$,
note that we do not see a similar behaviour toward other W--R stars in
the MUSE data.

Recent studies have demonstrated that the extinction law in 30 Doradus
is anomalous
\citep{2013A&A...558A.134D,2014A&A...564A..63M,2014MNRAS.445...93D},
with $R_V$\,$\approx$\,4.4, indicating higher total extinction and with
implications for nebular diagnostics.  Recalculating for
$R_V$\,$=$\,4.4 and an extinction law of the form from
\citet{2014A&A...564A..63M}, we find an average $c$(H$\beta$) of
$\approx$0.60\,mag. Given our use of relatively  red diagnostic lines
(and with small separation in terms of wavelength), this has a minimum
effect on our calculations of temperature and density in the following
section.

\subsection{Density and temperature distribution}\label{SECT:DENS}

\begin{figure*}[]
\resizebox{\hsize}{!}{\includegraphics[angle=0,width=\textwidth]{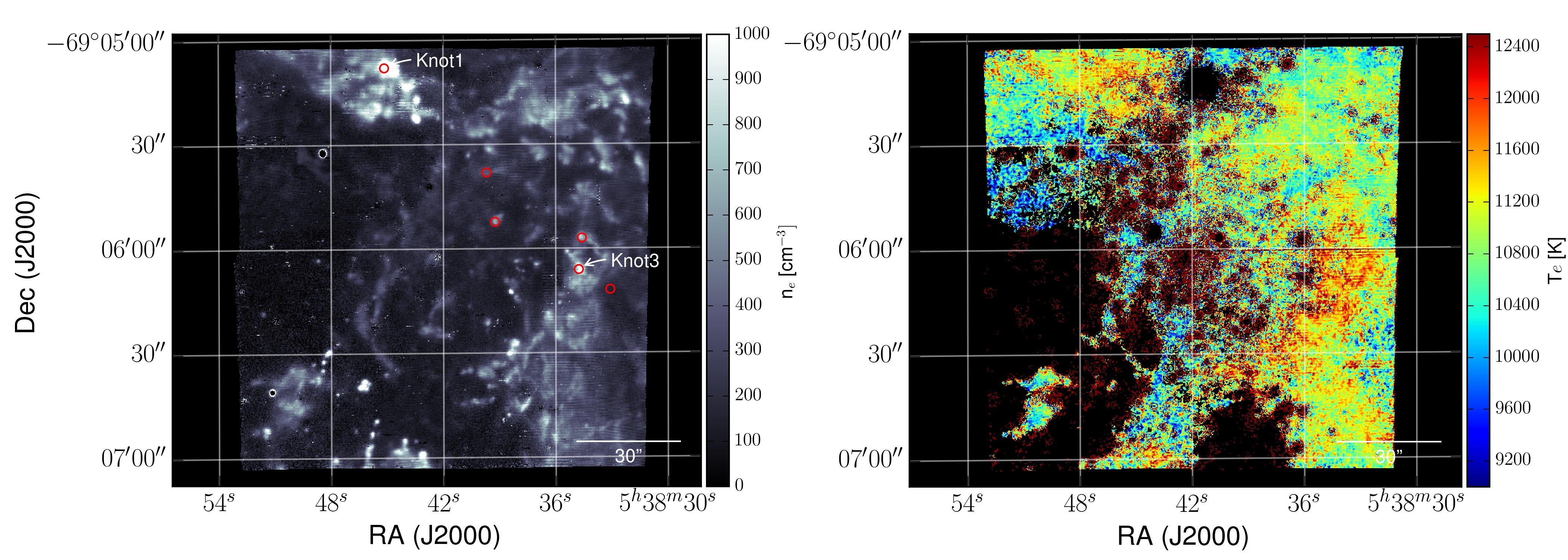}}
  \caption{{\it Left:} Electron density map derived from the
    [\ion{S}{ii}] $\lambda$6717/6731 ratio.  Candidate young stellar
    objects from \cite{2009ApJS..184..172G} are indicated by the red
    open circles, and Knots 1 and 3 from \cite{2002AJ....124.1601W}
    are also identified. {\it Right:} Electron temperature map derived
    from the [\ion{N}{ii}] lines.}
        \label{Fig:TEMP} \end{figure*}

The nebular lines observed in our MUSE datacube (see
Fig.~\ref{Fig:spec_neb}) can be used to construct maps of the electron
density and temperature \citep[e.g.][]{2015MNRAS.450.1057M}. The
electron density was estimated from the
[\ion{S}{ii}]\,$\lambda$6717/6731 ratio following the parameterisation
of \citet{1984MNRAS.208..253M}, with the average ratio implying a mean
density of $\approx$230 cm$^{-3}$. The left-hand panel of
Fig.~\ref{Fig:TEMP} presents the range of densities in the MUSE data,
spanning 50--1000 cm$^{-3}$, which agrees reasonably well with
\citet{2010ApJS..191..160P}.  The densest regions correspond to sites
of on-going stellar formation and high densities of CO
\citep{1998A&A...331..857J}. There are several dense regions to the
south, mainly in the tip of structures that resemble the pillars in
M\,16.

Figure~\ref{Fig:TEMP} also indicates the six sources in the MUSE mosaic
classified as `definite', `probable', or `possible' young stellar
objects (YSOs) by \cite{2009ApJS..184..172G}. Four of these coincide
with relatively dense regions: \textit{Knot~1}
\citep{1997ApJS..112..457W}, \textit{Knot~3} \citep{1999AJ....117..225W},
and a bullet-shaped clump around the YSO candidate close to R134 (cf.
Fig.~\ref{Fig:EXT}) that might be shaped by the powerful stellar wind
from this WN6 star.

The absence of [\ion{O}{iii}]\,$\lambda$4363 from the MUSE dataset led
us to use the [\ion{N}{ii}]$\lambda\lambda$\,6548,6584/5755 and [\ion{S}{iii}]$\lambda\lambda$\,9069/6312 
ratios (after correction for extinction) to estimate the electron
temperature, using the calibration of \citet{2006agna.book.....O}.
The right-hand panel of Fig.~\ref{Fig:TEMP} presents the electron
temperature map of NGC\,2070 based on the [\ion{N}{ii}] line ratios, which 
range from 9000 to 12500\,K, in
agreement with the range estimated by \citet{2010ApJS..191..160P}
using the [\ion{O}{iii}]$\lambda\lambda$4959,5007/4363 ratio.  The
average intensity ratio of [\ion{N}{ii}]$\lambda\lambda$\,6548,6584/5755 gave
a mean temperature of $\approx$11000\,K. The weakness or absence of
[\ion{N}{ii}]\,$\lambda$5755 prevented estimates in some areas (black
patches in the map); the noisy temperature distribution
arises from the general weakness of [\ion{N}{ii}]\,$\lambda$ 5755.
High electron temperatures are associated with the densest regions to
the north, west, and south.  
The [\ion{S}{iii}]$\lambda$\,9069/6312 
intensity ratio provided an average temperature of $\approx$9800\,K.

%%%%%%%%%%%%

\section{Kinematics} \label{VELO}

\subsection{Nebular velocity distribution}\label{SEC:KINO}

\begin{figure*}[]
\resizebox{\hsize}{!}{\includegraphics[angle=0,width=\textwidth]{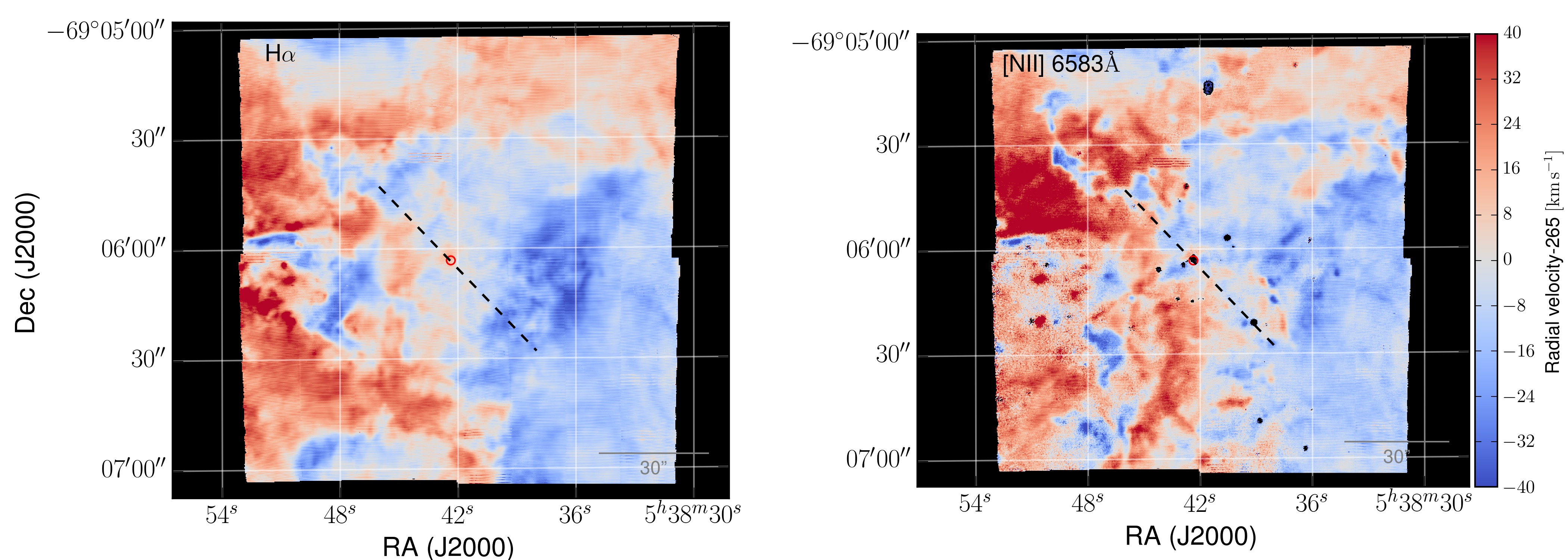}}
        
  \caption{Radial-velocity maps from Gaussian fits to the H$\alpha$
    and [N\,{\scriptsize{II}}]\,$\lambda$6583 emission (left and
    right panels, respectively). The core of R136 is indicated
with the red
    open circle, and the black dashed line is the proposed rotational
    axis from \cite{2012A&A...545L...1H}, see Sect.~\ref{SEC:KINO}.}
\label{Fig:GAUS_vel} \end{figure*}

The mean radial velocity estimated from our single-component Gaussian
fits to the H$\alpha$ and [\ion{N}{ii}] nebular lines was
265$\,$km\,s$^{-1}$ \citep[cf. 266\,$\pm$\,8\,\kms\
from][]{2012A&A...546A..73H}. We adopted this as our systemic value,
and differential velocities across the MUSE mosaic are shown for both
lines in Fig.~\ref{Fig:GAUS_vel}.

The estimated (absolute) velocities for the emission lines in
each of the extracted MUSE sources are listed in Cols. 8 and 9
of Table~\ref{TAB:cat}. These are effectively the average of the
peak emission in the pixels across each $\approx$1$''$ radius
aperture, and we estimate a pixel-to-pixel accuracy of
1.5--2\,\kms \citep[see also][]{2015A&A...582A.114W}. The nebular velocities for some sources have larger
dispersions, which probably are spatially resolved velocity 
structures within the extracted apertures and not poor precision (see
the complexity of the maps in Fig.~\ref{Fig:GAUS_vel}).

Some areas in the MUSE mosaic have differential velocities greater
than the range shown in Fig.~\ref{Fig:GAUS_vel} (i.e. |$\delta$v|
$>$\,40\,\kms), but we adopted the intensity scale to highlight the
general distribution. As noted by \cite{2010ApJS..191..160P}, there are
well-defined regions of both blue- and red-shifted material in
Fig.~\ref{Fig:GAUS_vel}, but the details and spatial sampling of MUSE
are striking. The ionised gas to the west of R136 is approaching us,
while the gas to the north-east of R136 is mostly receding, as noted by
\cite{1994ApJ...425..720C}. We also find a good match with the 
\cite{2017MNRAS.469.3424M} velocity measurements 
despite their higher spectral resolution.

The highest blue-shifted velocities roughly shape the shell in the
H$\alpha$ velocity map with peaks of $\approx$\,$-$35\,km\,s$^{-1}$
(and $\approx\,$$-$45\,km\,s$^{-1}$ for [\ion{N}{ii}]).
\textit{Knot~3} from \citet{1999AJ....117..225W}, labelled in
Fig.~\ref{Fig:TEMP}, is located in one of the most strongly blue-shifted regions.
Radiation pressure, stellar winds, and/or former supernovae could have
compressed the material producing this large shell structure
\citep[e.g.][and references therein]{2014MNRAS.442.2701R}. Assuming a
symmetric phenomenon, one might anticipate blue-shifted material
elsewhere around R136 (although projection effects should be borne in
mind), but this is not the case. The dense gas in the north-eastern
region has intense H$\alpha$ and [\ion{N}{ii}] emission
(Fig.~\ref{Fig:GAUS_INT}), but with a similar radial velocity as the
core of R136. Gas in the eastern cavity is predominantly red-shifted by
25-40\,km\,s$^{-1}$.

\begin{figure*}
        
\resizebox{\hsize}{!}{\includegraphics[angle=0,width=\textwidth]{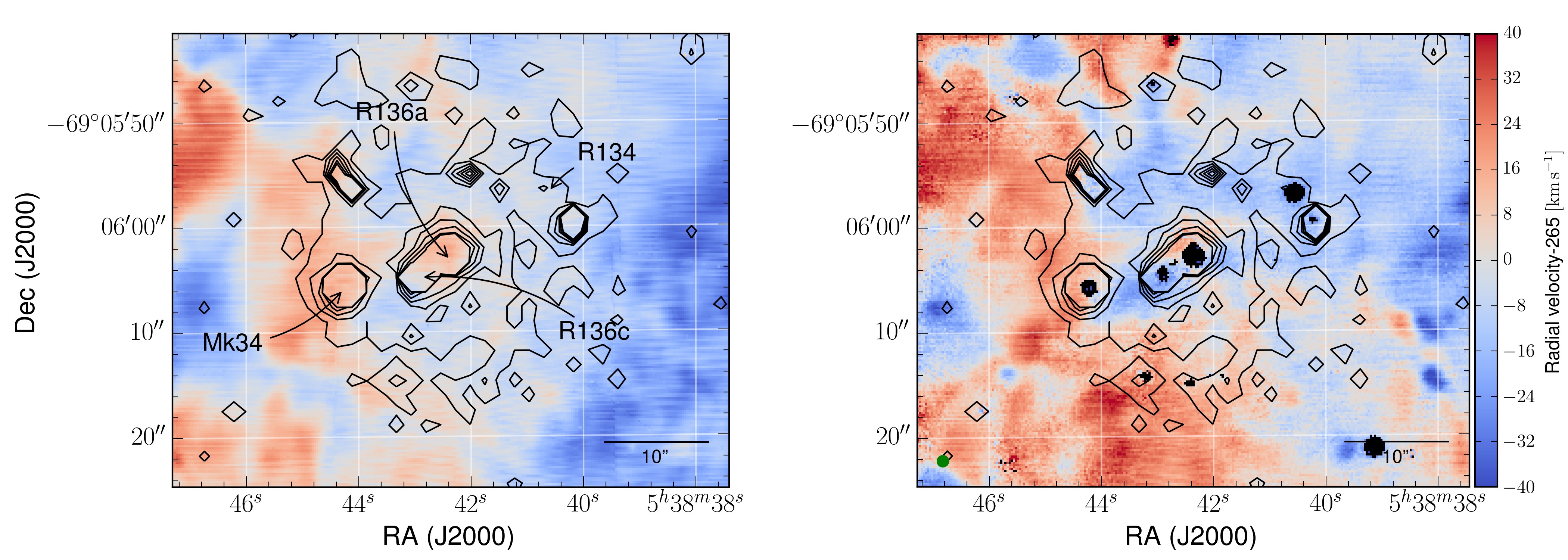}}

 \caption{Radial-velocity estimates for the peak H$\alpha$ and
    [\ion{N}{II}] emission (left- and right-hand panels, respectively)
    in the vicinity of R136 at the centre of the MUSE mosaic.
    Contours mark the most prominent X-ray sources from {\em Chandra}
    1.1-2.3\,KeV imaging
    \citep{2006AJ....131.2140T,2006AJ....131.2164T}, with the
    strongest sources labelled in the left-hand panel. Velocities are
    plotted on the same scale as in Fig.~\ref{Fig:GAUS_vel}.
    The average PSF of the MUSE data ($\approx$1$''$) is
    indicated in the right-hand panel by the green dot in the
    south-eastern corner.}
\label{Fig:XRAY_cent} \end{figure*}

Beyond \textit{Knot~3,} there are several other features of note in
the western shell.  There are two sources at approximately
$\alpha$\,$=$\,5$^{\rm h}$38$^{\rm m}$39$^{\rm s}$ and
$\delta$\,$=$\,$-$69$^\circ$06$'$10$''$ in the [\ion{N}{ii}] map (see
the western edge of the right-hand panel of Fig.~\ref{Fig:XRAY_cent}).
Bi-modal red- and blue-shifted material is seen, which is suggestive of either
rotation or ejecta, with velocities of $\approx$13 to $-$40\,\kms\ for
the northern feature. Given the ongoing star formation in NGC\,2070,
such phenomena are not unexpected \citep[e.g.][]{2015MNRAS.450.1057M},
although we note that while they are nearby, they are not part of the brightest
stellar nurseries in the western shell that we mentioned in Sect.~\ref{EMI}.

Red-shifted velocities are predominantly located on the south-eastern side of
R136 and in the optical cavity in the ISM on the eastern side of the
MUSE mosaic. In an optically thin environment (see
Sec.~\ref{SECT:EXT}), we could be observing background layers of 30~Dor
where the material is receding, but the cavity does not show a
homogeneous recession velocity. Only a portion is receding
(although some parts recede at more than 50\,\kms), with some of it approaching
us.

High velocities are found in the H$\alpha$ map around  several sources in the cavity,
which is clearly seen in the [\ion{N}{ii}] velocity map. As highlighted in Sect.~\ref{SIG}, this
coincides with a large FWHM in the emission as well, and unresolved
components probably play a role \citep[cf.
e.g.][]{2013A&A...555A..60T}. \cite{2017MNRAS.469.3424M} also reported
high-velocity clouds in the cavity. Broad and/or multiple components
are probably influencing our estimated velocities for this and other
objects, and high-resolution spectroscopy is required to
separate the different kinematic structures.

The overall spatial distribution of radial velocities is intriguing
and poses the question of whether localised independent phenomena or
a single mechanism dominate the behaviour. The largest red- and
blue-shifted velocities are qualitatively aligned with the core of
R136, along an axis almost perpendicular to the western shell;
a rotation of the ISM around this axis would induce a bi-modal radial-velocity distribution. There is marginal evidence to date for rotation
in young clusters
\citep[e.g.][]{1993AJ....105..938F,2011MNRAS.411.1386D}, but we note
the finding of \citet{2012A&A...545L...1H} of potential rotation of
the O-type stars of R136 with an amplitude of $\approx$3\,\kms. This is
much lower than the apparent difference in ISM velocities across R136,
although not a like-for-like comparison, as the ISM analysis extends
farther out into NGC\,2070. Nonetheless, the proposed rotational axis
from \cite{2012A&A...545L...1H} appears somewhat offset from that
suggested by our ISM measurements (see Fig.~\ref{Fig:GAUS_vel}).

External factors acting on NGC 2070, combined with the outward
pressure from the young stellar population of R136, might offer an
alternative mechanism. \cite{2012A&A...540L..11S} showed that most
star clusters form at the junction of filaments in molecular clouds, and
\cite{2014prpl.conf...27A} reviewed the evidence from theory and
observations that converging filaments provide sufficient mass flux
onto the filament junction to allow a star cluster to form.  In this
paradigm of massive cluster formation, we expect 30 Dor to be fed by a
number of accreting filaments of molecular gas. In the time since the
cluster has formed, feedback processes (winds and supernovae) have
created a hot bubble of high-pressure, X-ray emitting gas around R136,
and this prevents further accretion of dense gas onto the cluster
\citep[e.g.][]{2011ApJ...738...34P}. R136 is at the centre of
a young superbubble that is being formed by feedback processes from its stars
\citep{1994ApJ...425..720C}.

The result is that the star cluster could be embedded in a stream of dense
gas that is forced to flow around it, a well-known physical phenomenon
explored at different astropysical scales
\citep[e.g.][]{1982A&A...110..300S,2015A&A...573A..10M}. The ISM
may collide with the hot bubble surrounding R136 and encircle the
cluster, producing a ram-pressure effect and moving the ISM toward the
observer, consistent with production of the western shell.

Furthermore, the feedback from massive stars in R136 has swept up a
massive  (65\,000 M$_\odot$, \citealt{1994ApJ...425..720C}) dense shell that is expanding outwards at
up to 40\,\kms.  This expanding shell would decelerate and entrain any
ISM that would be flowing towards R136.  The observed eastern cavity may be
a lower-density part of the ISM, where there were no accreting
filaments.  Under this hypothesis,  the mean flow of gas would then be from
west to east, and the cavity is a low-density wake downstream from
R136.  The dent in the CO maps around R136 reported by \citet[][see their Fig.~1, for instance,]{1998A&A...331..857J} may be a consequence of the
strong UV radiation as they suggested, but would also support the idea
of differential movement between R136 and the ISM.

Moreover, the eastern cavity has much lower extinction than the western
side, suggesting that the expanding shell around R136 is incomplete in
the hemisphere closer to us. The H$\alpha$ emitting gas would then be
mainly emitted from the far side of the shell, explaining why it is strongly
red-shifted.  This hypothetical scenario will be tested in future works.

\subsection{Gas kinematics around R136} \label{VELO_core}

In addition to R140, the main X-ray sources in the MUSE mosaic are in
R136 \citep{2006AJ....131.2140T,2006AJ....131.2164T}. The brightest
X-ray sources are Mk\,34 (WN5h), R136c (WN5h), and the R136a cluster
\citep{2009MNRAS.397.2049S}, presumably involving colliding-wind
binaries \citep{2010MNRAS.408..731C}. Periodic X-ray
variability in Mk\,34 consistent with a colliding-wind system has
recently been discovered from observations with {\em Chandra} as part
of the T-ReX project \citep{doi:10.1093/mnras/stx2879}.
Figure~\ref{Fig:XRAY_cent} shows the central
$\approx$40$''$\,$\times$\,40$''$ ($\simeq$10\,$\times$\,10\,pc) of the
H$\alpha$ and [\ion{N}{ii}] velocity maps.  Red-shifted material
almost encircles the X-ray detections, which is particularly prominent on
the eastern side of the cluster.  The projected red-shifted ISM component, 
observed in Fig.~\ref{Fig:GAUS_vel}, extends from the east side of the field 
to the core of R136. Blue-shifted material around the core of R136 may trace escape
channels for the radiation pressure. Although the FWHM map does not
appear to show blended clouds (see Fig.~\ref{Fig:GAUS_INT}) and/or largely broad profiles around the
core (as it does in the cavity), a composite spectrum is expected in
the cluster core \citep{1999MNRAS.302..677M}, but will be unresolved
in the MUSE data.

We would expect the rich population of massive stars in
R136 
to be driving material away as a result of their combined intense stellar
winds.  The ISM
apparently moves toward the core of R136 (cf. the systemic velocity of
the cluster). However, the projected red-shifted material on R136a could 
be either a foreground/background ISM projection, or it might
mean that the radial velocity of
R136a is higher than the average systemic velocity of NGC\,2070.

\begin{figure*}[] 
\resizebox{\hsize}{!}{\includegraphics[angle=0,width=\textwidth]{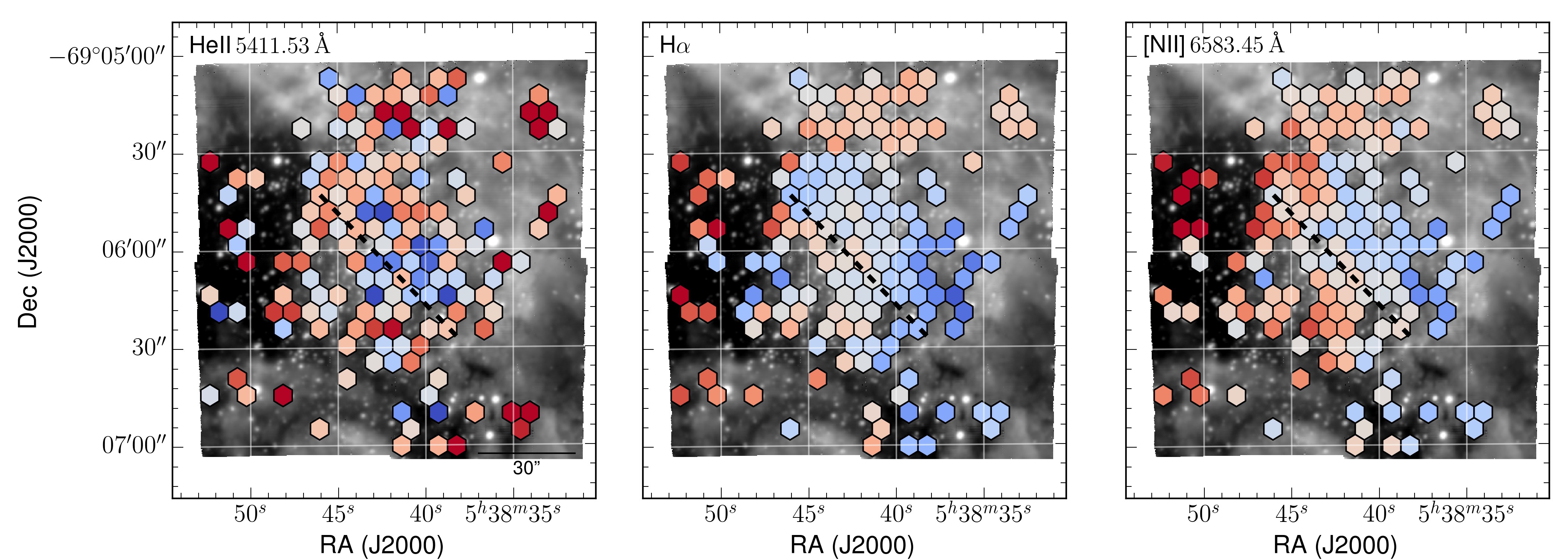}}
  \caption{Radial-velocity estimates from
    \ion{He}{ii}\,$\lambda$5411.5 (left-hand panel) for 270 early-type
    stars compared with estimates from the H$\alpha$ and
    \ion{N}{ii}\,$\lambda$6583.45 nebular lines superimposed on the
    spectra (central and right-hand panels, respectively).  The colour
    of each hexagon indicates the median velocity of the sources in
    that area  and follows the same
    scale as in Fig.~\ref{Fig:GAUS_vel}.  The black dashed line in
    each panel is the rotational axis proposed by
    \cite{2012A&A...545L...1H}.}\label{Fig:STARS} \end{figure*}

\subsection{Stellar velocity distribution} \label{STAR}

To investigate stellar radial velocities, we used the
\ion{He}{ii}\,$\lambda$5411.5 line in the O-type spectra (and some of
the earliest B-type spectra, see Fig.~\ref{Fig:spec}); using the
\ion{He}{ii} line has the advantage that it is free of potential nebular
contamination; compare  the \ion{He}{i} lines in the more numerous B-type
stars. Estimated radial velocities are given in Table~\ref{TAB:cat}
for 270 stars, with an average of 271\,$\pm$\,41\,\kms\ \citep[in
agreement with the systemic value of the ISM and stellar results
from][]{2012A&A...546A..73H}. The uncertainties quoted on the
velocities in Table~\ref{TAB:cat} are the standard deviations of each
pixel within the extracted aperture.

From an analysis of 38 (apparently non-variable) O-type stars,
\cite{2012A&A...546A..73H} concluded that the core of R136 was in
virial equilibrium with a line-of-sight velocity dispersion of no more
than 6\,\kms.  In this context, the large dispersion of 41\,\kms\ from
our estimates probably reflects the limits of the velocity resolution
from MUSE (and the S/N of our spectra), combined with undetected
binaries \citep[see][]{2009AJ....137.3437B,2012A&A...546A..73H}. The
distribution of stellar radial velocities is shown in
Fig.~\ref{Fig:STARS}, compared with estimates from the nebular
H$\alpha$ and [\ion{N}{ii}] at the same positions. The large
dispersion potentially arising from undetected binaries (and the lower
S/N of the \ion{He}{ii} absorption; cf. the strong emission lines)
precludes an obvious link between the stellar and gas components.

The (spatially limited) view of the ISM dynamics in
Fig.~\ref{Fig:STARS} agrees reasonably well with the rotational axis
proposed by \cite{2012A&A...545L...1H}.  However, as noted earlier
from consideration of the full spatial information
(Fig.~\ref{Fig:GAUS_vel}), the true picture with regard to potential
rotation of R136 is likely more complicated.  Somewhat surprisingly, we
note that the direction of the  velocity gradient of the gas in
Fig.~\ref{Fig:STARS} is inverted; compare the stellar results from
\citeauthor{2012A&A...545L...1H} (gas red-shifted more in the
 south-east, stars more so in the north-west).  In short, although the
ISM and stellar population have comparable average velocities in
space, the potentially differential movement of these components could add
support to the hypothesis that external mechanisms shape the dynamics
of the ISM.

\section{Summary} \label{CONC}

We introduced VLT-MUSE observations of the central 2$'$$\times$2$'$
of NGC\,2070 within the Tarantula Nebula in the LMC. The data permit
a homogeneous spectroscopic census of the massive stars and ionised
gas in the vicinity of the central star cluster R136.

We provide a catalogue of 2255 point sources detected in the MUSE
mosaic, and include $V$- and $I$-band magnitudes estimated from the
datacubes.  The colour-magnitude diagram reveals a young bright
population and a fainter group of stars, with the latter matching the
predicted location of pre-main-sequence stars
\citep[e.g.][]{2012MNRAS.427..127B}.  Both populations are compatible with a
recent burst of star formation spanning a few Myr, as proposed in
recent studies
\citep[e.g.][]{2015ApJ...811...76C,2016MNRAS.458..624C}. We also
revisited the known W--R population in the region and their fluxes in
some of the most prominent emission lines.

We constructed the extinction map for NGC\,2070 from the ratio of
H$\alpha$/H$\beta$, a map of electron densities based on [\ion{S}{ii}] lines, and
 the electronic temperature distribution maps from relative ratios of the [\ion{N}{ii}] and [\ion{S}{iii}] lines. The
average electron density (230\,cm$^{-3}$)  and temperature (11000\,K and 9800\,K, 
based on  [\ion{N}{ii}] and [\ion{S}{iii}], respectively) agree with those from \cite{2010ApJS..191..160P}, but
the spatial resolution provided by MUSE gives new insights into the
structure of the ISM and ongoing star formation in NGC\,2070. We
resolved several high-density clumps that are probably linked to the
formation of new stars, some of which have previously been classified
as candidate young stellar objects
\citep{2002AJ....124.1601W,2009ApJS..184..172G}. The electron
temperature map shows higher temperatures close to these dense areas.

The structural features in the ISM of the region were investigated
using Gaussian fits to the H$\alpha$ and [\ion{N}{ii}]\,$\lambda$6583
nebular emission lines.  The H$\alpha$ emission traces several known
structures in NGC\,2070: a prominent shell in the west, several
filaments in the north of the field, and a cavity to the east of R136.
The extinction map resembles these structures, but we find several
areas of high extinction in the south-west of the field without
counterparts in the H$\alpha$ map. The [\ion{N}{ii}] emission map
shows a more clumpy distribution (similar to the high-density areas
traced by the electron-density map) and is probably related to active
sites of star formation.

We estimate a systemic velocity for the ISM of 265\,\kms\ (in
agreement with previous studies), but the velocity maps for H$\alpha$
and [\ion{N}{ii}] show complex kinematics in the region. We find a
bi-modal, blue- and red-shifted distribution in the gas velocities
centred on R136; the western shell is mainly blue-shifted, with 
material in the eastern cavity receding
\citep[e.g.][]{1994ApJ...425..720C}. This bi-modality does not match
the velocities and rotational axis for stars in R136 from
\citet{2012A&A...545L...1H}, but differential velocities between the
stellar cluster and the ISM might arise from the effects of ram
pressure. 

Within our velocity maps we find several interesting point sources
that might be related to stellar runaways, jets, the formation of new
stars, or interaction of the gas with the W--R stars, and warrant
further study. Closer inspection of the central region reveals
red-shifted material encircling the diffuse X-ray emission, and
reaching as far as R136a in the core.  This could point to a 
more complex kinematics that are unresolved in these data.  Several
blue-shifted clumps around R136 suggest escape channels for the hot
gas and radiation.

We estimate radial velocities of 270 O-type stars, finding an average of
271\,$\pm$\,41\,\kms. The stellar velocities broadly match those of
the ISM, but the spectral resolution of the data and likely presence
of undetected binary components limits further analysis of the stellar
dynamics.

Future papers on these data will focus on the stellar content of
NGC\,2070 and its integrated stellar and nebular properties \citep{2017Msngr.X...XC}.  Given
that the MUSE mosaic would only subtend 0.6$''$ at a distance of
10\,Mpc (comparable to long-slit spectroscopy of extragalactic
\ion{H}{ii} regions with large ground-based telescopes), the
integrated properties will be of interest as they provide a template with
which to investigate extragalactic systems.

%Added by TeX Support
\longtab{
\begin{longtable}{rrrrrrrrrr}
\caption{\label{TAB:cat} Observational catalogue for extracted MUSE sources in NGC\,2070. Cols. 1, 2, 3: identifications in current work and
          from \cite{1999A&A...341...98S} and \cite{2011A&A...530A.108E},
          respectively. Cols. 6, 7: $V$-band magnitudes and $(V-I)$ colours estimated from
          the MUSE data,
          although see Sect.~\ref{phot} for calibration of zero-points. Cols.
          8, 9: estimated radial velocities from (single-component) fits to the H$\alpha$ and
          [\ion{N}{ii}]\,$\lambda$6583.45 nebular emission lines. Col. 10:
          estimated radial velocity from stellar \ion{He}{ii}\,$\lambda$5411.5
          absorption (see Sect.~\ref{STAR}).}\\
\hline\hline
        MUSE ID & Selman ID & VFTS ID & $\alpha$ (J2000) & $\delta$ (J2000) & $V_{\rm MUSE}$ & ($V-I$)$_{\rm MUSE}$ & RV (H$\alpha$) & RV ([\ion{N}{ii}]) & RV (\ion{He}{ii})\\
          & & & [${\rm h,m,s}$] & [${\circ,',''}]$ & [mag] & [mag] & [\kms] & [\kms] & [\kms] \\
\hline
\endfirsthead
\caption{continued.}\\
\hline\hline
       MUSE ID & Selman ID & VFTS ID & $\alpha$ (J2000) & $\delta$ (J2000) & $V_{\rm MUSE}$ & ($V-I$)$_{\rm MUSE}$ & RV (H$\alpha$) & RV ([\ion{N}{ii}]) & RV (\ion{He}{ii})\\
          & & & [${\rm h,m,s}$] & [${\circ,',''}]$ & [mag] & [mag] & [\kms] & [\kms] & [\kms] \\
\hline
\endhead
\hline
\endfoot
2765 & -- & -- & 5:38:31.368 & -69:05:34.44 & 19.67 & -0.01 & 278$\pm$3.7 & 276$\pm$6.4 & -- \\
2363 & 1087 & -- & 5:38:31.392 & -69:05:18.96 & 18.92 & 0.22 & 284$\pm$7.1 & 284$\pm$1.4 & -- \\
2217 & -- & -- & 5:38:31.392 & -69:05:21.84 & 21.7 & 0.69 & 280$\pm$4.2 & 278$\pm$1.4 & -- \\
1706 & -- & -- & 5:38:31.392 & -69:06:01.8 & 21.24 & 0.88 & 254$\pm$1.5 & 250$\pm$1.6 & -- \\
413 & -- & -- & 5:38:31.416 & -69:06:52.56 & 19.72 & 0.81 & 255$\pm$2.0 & 252$\pm$1.8 & -- \\
2090 & -- & -- & 5:38:31.416 & -69:05:53.16 & 16.45 & 0.4 & 256$\pm$2.1 & 248$\pm$2.0 & -- \\
1968 & 989 & -- & 5:38:31.416 & -69:05:57.12 & 16.77 & 0.12 & 254$\pm$1.8 & 251$\pm$1.7 & -- \\
2485 & -- & -- & 5:38:31.416 & -69:05:20.04 & 21.76 & 1.56 & 280$\pm$6.8 & 280$\pm$2.3 & -- \\
\end{longtable}
}

%-----------------------------------------

%############################################################

\begin{acknowledgements}
The authors thank the referee for useful comments and helpful suggestions that improved this manuscript.
We would like to thank S. Yeh for kindly providing near-IR images taken with the NEWFIRM camera on the 
Blanco telescope on Cerro Tololo (Chile). We thank Anna McLeod for her thoughts on the manuscript 
and ideas for future directions. JM acknowledges funding from a Royal Society--Science Foundation Ireland University Research Fellowship. This research made use of Astropy, a community-developed core Python package for Astronomy \citep{2013A&A...558A..33A}, and APLpy, an open-source plotting package for Python \citep{2012ascl.soft08017R}.

\end{acknowledgements}

%-------------------------------------------------------------------

%############################################################ % BIBLIOGRAPHY 
%############################################################

\bibliographystyle{aa}

\bibliography{AA_2017_32084_BIB}

%############################################################ % Appendix 
%############################################################

\end{document}